\documentclass[twoside,nofootinbib,preprintnumbers,aps,11pt,longbibliography,notitlepage]{revtex4-1}
\usepackage{euscript,amssymb}
\usepackage{amsthm}
\usepackage{graphicx}
\usepackage{amsfonts}
\usepackage{amsmath}
\usepackage{amssymb}
\usepackage{fancyhdr}
\usepackage{epsfig}
\usepackage{color}
\usepackage{esint}
\usepackage{soul}
\usepackage{calc}
\usepackage{enumerate}
\usepackage{afterpage}
\usepackage[papersize={8.5in,11in}]{geometry}

\allowdisplaybreaks

\newcommand{\be}{\begin{equation}}
\newcommand{\ee}{\end{equation}}

\begin{document}

\title{The Petrov type D isolated null surfaces.}

\author{Denis Dobkowski-Ry{\l}ko}
	\email{Denis.Dobkowski-Rylko@fuw.edu.pl}
	\affiliation{Faculty of Physics, University of Warsaw, ul. Pasteura 5, 02-093 Warsaw, Poland}
\author{Jerzy Lewandowski}
	\email{Jerzy.Lewandowski@fuw.edu.pl}
	\affiliation{Faculty of Physics, University of Warsaw, ul. Pasteura 5, 02-093 Warsaw, Poland}
\author{Tomasz Paw{\l}owski}
	\email{Tomasz.Pawlowski@fuw.edu.pl}
	\affiliation{Center for Theoretical Physics, Polish Academy of Sciences, Al. Lotnik\'ow 32/46, 02-668 Warsaw, Poland.}
	\affiliation{Faculty of Physics, University of Warsaw, ul. Pasteura 5, 02-093 Warsaw, Poland}

\begin{abstract}    
Generic black holes in vacuum-de Sitter / Anti-de Sitter spacetimes are studied in quasi-local framework, where the relevant
properties are captured in the intrinsic geometry of the null surface (the horizon). Imposing the quasi-local notion of
stationarity (null symmetry of the metric up to second order at the horizon only) we perform the complete classification
of all the so-called \emph{special Petrov types} of these surfaces defined by the properties (structure of principal null direction)
of the Weyl tensor at the surface. The only possible types are: II, D and O. In particular all the geometries of type O are identified.
The condition distinguishing type D horizons, taking the form of a second order differential equation on certain complex invariant
constructed from the Gaussian curvature and the rotation scalar, is shown to be an integrability condition for the so-called near
horizon geometry equation. The emergence of the near horizon geometry in this context is equivalent to the  hyper-suface orthogonality
of both double principal null directions. We further formulate a no-hair theorem for the Petrov type D axisymmetric null surfaces
of topologically spherical sections, showing that the space of solutions is uniquely parametrized by the horizon area and angular momentum.
\end{abstract}

\date{\today}

\pacs{???}

\maketitle
\section{Introduction}
The theory of isolated null surfaces  is a part of the local approach to  black holes (BH) \cite{ABL1,ABL2,AK}. 
 Intrinsic geometry of those surfaces, that is the induced  metric  and the induced  covariant derivative,   
has physically relevant features of the  geometry of stationary  BH horizons. The theory is applicable to cosmological 
horizons and to the null boundaries of the conformally compactified    asymptotically flat spacetimes \cite{AshtekarMagnon}. 
Moreover, it may be applied to the black hole holograph construction of spacetimes about Killing horizons \cite{R1, R2}.  
There are analogies between the properties   of isolated null surfaces  and the properties of BHs.
Isolated null surfaces  admit their  mechanics, an analog of the BH  "thermodynamics" \cite{ABL2},   
the rigidity theorem \cite{LPsymmetric}, 
and the uniqueness theorems \cite{LPextremal, LPkerr}. 
 The key difference between the BHs and the isolated null surfaces lies  in the degrees of freedom.  While the families
 of stationary BH solutions are finite dimensional, the intrinsic geometry   of isolated null surfaces has  local  degrees 
 of freedom. That makes their theory far more general.  
 The current paper develops  the observation that   at each non-extremal isolated null surface  the spacetime   Weyl tensor 
 can be  determined  by the  intrinsic geometry via the  Einstein equations and some stationarity assumption. 
 We find all the possible Petrov types of isolated null surface and focus on the case of the Petrov type D.  
 We derive and investigate the Petrov type D equation on data defined on a $2$-slice of an isolated null surface: a Riemannian metric
  $g_{AB}$ and a co-vector $\omega_A$ modulo gradient.  Moreover, we find all the Petrov type O geometries. We show,  that the Petrov type D equation is also an integrability condition 
  for the so-called near horizon geometry equation \cite{NHG}, first discovered and studied as the extremal isolated horizon equation 
  \cite{ABL1,LPextremal}.  We find the geometric consequences  of the emergence of the near horizon 
  geometry  equation in that new context. It is   non-twisting of the double principal null directions of the Weyl tensor.    
 We also characterize type D isolated null surfaces that have spherical section and are axisymmetric. 
 A detailed derivation of those  solutions is contained in the accompanying paper \cite{DLP2}. The result is
 a generalization of the earlier geometric result \cite{LPkerr} to the case of non-zero cosmological constant. 
 
Physically, the results presented in this paper are relevant for the issue of the local, geometric characterization of the    
horizons contained in the Kerr $-$ (anti) de Sitter  spacetimes. They answer the question of what local properties 
single out the Kerr $-$ (anti) de Sitter horizons in a much larger class of all the isolated null surfaces. The characterization
of the axisymmetric solutions we formulate can be considered as a local no hair theorem. 

Geometrically,  the Petrov type D equation is a new equation of mathematical physics, that
seems to  lead to new interesting structures that can be considered in a $2$-dimensional Riemannian 
geometry.   Also, as an integrability condition for the near horizon geometry equation, the Petrov  type D equation 
may be useful for solving the problem of dimensionality of the space of solutions to the near horizon geometry equation.

\section{Isolated null surfaces} \label{NE-SF}
\subsection{Notation and convention}
Throughout this paper we consider  a $4$-dimensional  spacetime that consist of a manifold  $M$  and a metric tensor $g_{\mu\nu}$ of the signature $-+++$.  By $\nabla_\mu$ we  denote the  torsion free  covariant derivative in $M$,  corresponding to $g_{\mu\nu}$ via

$$\nabla_\alpha g_{\mu\nu}\ =\ 0.$$   

About the metric tensor we  assume   the vacuum Einstein equations with a cosmological constant $\Lambda$,
\begin{equation}\label{EE} G_{\mu\nu} + \Lambda g_{\mu\nu}\ =\ 0, \end{equation}
where $G_{\mu\nu}$ is the Einstein tensor.  

In $M$ we study  a  $3$-dimensional null surface $$H\subset M.$$ 
We assume that $H$ contains a slice  that intersects each 
null curve in $H$ exactly once. In other words, $H$ has the topology
\begin{equation}\label{SxR} H \ =\ S\times \mathbb{R},\end{equation} 
where $S$  is the $2$-dimensional space of the null curves in $H$ assumed to be connected.  

Throughout this paper we use the following (abstract) index  notation \cite{Wald}:
\begin{itemize}
\item  Indices of the spacetime tensors  are denoted by
lower Greek letters: $ \alpha, \beta,\gamma, ... \ = 1,2,3,4$
\item  Tensors defined in $3$-dimensional  space $H$  carry indexes denoted by lower Latin letters: $a, b, c, ...\ = 1,2,3$
\item  Capital Latin letters $A, B, C, ... \ = 1,2$ are used as the indexes of tensors defined in $S$.  
\end{itemize}

Non-expanding and isolated null surfaces in four dimensions can be dealt with by either the covariant framework  or  using the 
Newman-Penrose (NP) formalism  \cite{NP}.  Both of the approaches are used in \cite{ABL1}: the covariant one in the main part, and the NP one 
in Appendix B of that work.  In the current paper  we use  the covariant framework as an official language. 
Of course, in some derivations we apply adapted null frames. Then, locally in the paper,  we introduce necessary elements of the NP  formalism explicitly
on a current basis. Our notation is consistent with that of \cite{ABL1} and \cite{NP}, except that as in \cite{ABL1}, the vectors $k$, $\ell$ introduced 
in \cite{NP} correspond to vectors $\ell$, $n$ in our work.

\subsection{Non-expanding shear-free null surfaces}
The spacetime metric tensor $g_{\mu\nu}$ induces a degenerate metric tensor $g_{ab}$ in the null surface $H$. 
The degeneracy means, that at every point  $x\in H$ there is a vector $\ell \in T_xH$
such that 
\begin{equation}\label{lh}
\ell^a g_{ab}\ =\ 0 .
\end{equation}
The distribution of the degenerate direction  is a $1$-dimensional sub-bundle $L\subset T(H)$. 
Every section $\ell \in \Gamma(L)$  of $L$ is a non-trivial vector field tangent to $H$ and orthogonal to $H$, including itself, at every point. 
The integral curves tangent to $L$ are null at every point, they foliate $H$ and each of them is  geodesic in the spacetime $ M$.  
We refer to them as null generators of $H$.  
\medskip

\noindent{\bf Assumption 0:}  {\it For every pair $X$ and $Y$ of vector fields tangent to $H$,  the spacetime 
vector field  $X^\alpha\nabla_\alpha Y^\mu$ is also tangent to $H$,}
\begin{equation}
X,Y\in \Gamma(T(H)) \ \ \Rightarrow \ \ \nabla_X Y \in \Gamma(T(H)) .  
\end{equation}
In other words,  the spacetime covariant derivative $\nabla_\mu$  preserves the tangent bundle 
$T(H)$ and endows $H$ with a  covariant derivative $\nabla_a$ via the restriction. 
\medskip 

An equivalent assumption would be  the vanishing of the extrinsic curvature of $H$,  that is 
\begin{equation}
X^\alpha Y^\beta \nabla_\alpha \ell_\beta=0
\end{equation}
for every pair $X,Y\in \Gamma(T(H))$ and every $\ell\in \Gamma(L)$.

\noindent{\bf Definition 0.} {\it The pair $(g_{ab}, \nabla_a)$  is called the intrinsic  geometry of $H$.}
\medskip
   
The derivative $\nabla_a$ in $H$ is torsion free and satisfies the pseudo metricity condition 
\begin{equation}\label{Dh}   \nabla_c g_{ab}\ =\ 0.  \end{equation}
It follows from those properties of $\nabla_a$ and from the degeneracy of $g_{ab}$ that  
for every $\ell \in\Gamma(L)$, the Lie derivative of $g_{ab}$ vanishes \cite{LPhigher}
\begin{equation}\label{Llh}
{\cal L}_\ell g_{ab}\ =\ 0 .
\end{equation}
 That property could be yet another equivalent version of Assumption 0.

If $g_{ab}$ were a degenerate metric tensor in a general null surface, we would decompose ${\cal L}_\ell g_{ab}$ uniquely  into the 
expansion $\theta$ and shear $\sigma_{ab}$
\begin{equation}\label{thetasigma}
\frac{1}{2}{\cal L}_\ell g_{ab}\ =\ \frac{1}{2}\theta g_{ab} + \sigma_{ab},
\end{equation}
where $\sigma_{ab}$ is traceless in each of the $2$-dimensional fibers of $T(H)/L$. Hence, our surface  $H$ 
(\ref{Llh}) is non-expanding and shear-free (NE-SF).  The opposite is also true: if a null surface is non-expanding 
and shear-free then the spacetime covariant derivative $\nabla_\alpha$ preserves the bundle $T(H)$ \cite{ABL1}. 
\medskip

\noindent{\bf Remark.} We have assumed that surface $H$ is both, non-expanding and shear-free.  
However,  whenever the equations (\ref{EE}) are satisfied,  the vanishing of the expansion $\theta$ 
implies  the vanishing of the shear $\sigma_{ab}$ (\ref{thetasigma}).  This is a standard application of the Raychaudhury equation
 that is satisfied by a geodesic $\ell\in \Gamma(L)$  (that is such that $\ell^a\nabla_a\ell=0$), namely
\begin{equation}
l^a(\theta)_{,a}\ =\ -\frac{1}{2}\theta^2 - ||\sigma ||^2 - R_{\mu\nu}\ell^{\mu\nu}. 
\end{equation}
The second term is the norm of the shear in the $2$-dimensional space of vectors orthogonal to $L$ modulo $L$, and the 
third vanishes owing to the vacuum Eintein equations (\ref{EE}) 
\cite{ABL1}. For that reason "non-expanding" sufficiently well characterizes the properties of $H$.  

\subsection{Ingredients of NE-SF null surface geometry $(g_{ab}, \nabla_a)$}
We continue the discussion of the intrinsic geometry $(g_{ab}, \nabla_a)$ of a  non-expanding and shear-free null surface $H$. 
Now we will use the natural projection  
\begin{equation}\label{Pi1} \Pi : H\rightarrow S.\end{equation} 
onto the space $S$ of the null curves in $H$. 
It follows from (\ref{lh}, \ref{Llh}) that on $S$ there is a uniquely defined Riemannian metric tensor $g_{AB}$,  such that
the degenerate metric $g_{ab}$ is the pullback by $\Pi^*$ of $g_{AB}$,
\begin{align}\label{Piq} g_{ab}\ =\ \Pi^*{}_{ab}{}^{AB}g_{AB}. \end{align}
In this way, the degenerate metric tensor  $g_{ab}$ in $H$ is all determined by the Riemanian metric $g_{AB}$ defined on the $2$ dimensional manifold $S$.  The metric $g_{AB}$ determines the unique torsion free covariant derivative $\nabla_A$ in $S$ that satisfies the metricity condition
$$\nabla_Ag_{BC}\ =\ 0.$$

The action of  the covariant derivative $\nabla_a$ in $H$ on a co-vector field $W\in\Gamma(T^*(H))$ orthogonal to the fibers of $L$, that is such that for every $\ell\in\Gamma(L)$
$$ \ell^aW_a\ =\ 0 ,$$
is determined by $g_{ab}$ \cite{AshtekarMagnon}.  Specifically
$$ \nabla_aW_b\ =\ \partial_{[a}W_{b]}\ +\ \frac{1}{2} {\cal L}_{\hat{W}} g_{ab}$$ 
where $\hat{W}^a$ is any vector field on $H$ such that
$$  \hat{W}^ag_{ab}\ =\ W_b. $$
We can think of that part of $\nabla_a$ as coinciding (vaguely speaking) with the covariant derivative $\nabla_A$ on $S$.
The remaining part of $\nabla_a$ is independent of $g_{ab}$. It  can be defined by introducing on $H$ any co-vector field $n_a\in \Gamma(T^*(H))$ that is nowhere orthogonal to $L$,
and considering the tensor 
\begin{equation}\label{Sab} S_{ab}\ :=\ \nabla_an_b.\end{equation} 
We may choose 
$$n_a:=-\nabla_av$$ 
and a function $v:H\rightarrow \mathbb{R}$ adapted to  a given non-vanishing vector field $\ell\in\Gamma(L)$,
such that
$$\ell^a\partial_a v =1.$$
Then, 
$$S_{ab}\ =\ S_{(ab)},$$ \
and 
$$\ell^an_a\ =\ -1, \ \ \ \ \ \ \ \ {\rm and} \ \ \ \ \ \ \ \  {\cal L}_\ell n_a\ = 0.$$ 

The one form
$$ \omega^{(\ell)}_a\ :=\ \ell^b S_{ab} $$
depends only on $\ell$ and may be defined by the equality
\begin{equation}\label{omega}
\nabla_a \ell^b\ =\ \omega^{(\ell)}_a\ell^b.   
\end{equation}
The function  
\begin{equation}\label{kappa}
 \kappa^{(\ell)}\ =\ \omega^{(\ell)}_a\ell^a, 
\end{equation}
is a self-acceleration of $\ell$,
\begin{equation}
\ell^a \nabla_a \ell^b\ =\  \kappa^{(\ell)}\ell^b. 
\end{equation}

While the rotation 1-form potential transforms with rescaling $\ell$,
\begin{equation}
\omega^{(f\ell)}_a = \omega^{(\ell)}_a + \partial_a{\rm ln}f,
\end{equation}
its external derivative 
\begin{equation}\label{Omega}
\Omega_{ab}\ =\  \partial_a\omega^{(\ell)}_b - \partial_b\omega^{(\ell)}_a.
\end{equation}
is  invariant  of the intrinsic geometry $(g_{ab}, \nabla_a)$ .
\medskip

\noindent{\bf Definition 1.} {\it  The $1$-form  $\omega^{(\ell)}$   (\ref{omega}) is called  rotation $1$-form potential, 
the function $\kappa^{(\ell)}$ (\ref{kappa}) is called  surface gravity, the $2$-form $\Omega$ (\ref{Omega}) is called the rotation 
$2$-form.}
\medskip

 The word {\it rotation}   may be understood just as a part of the definition. However,
when the angular momentum of a NE-SF null surface is well defined, then  it is indeed given by an integral of an 
expression proportional  to $\Omega$ \cite{ABL1}. 

\subsection{The vacuum constraints on the intrinsic geometry of the NE-SF null surfaces}
 The vacuum Einstein equations (\ref{EE}) with the cosmological constant $\Lambda$ imply 
 constraints on the  intrinsic geometry $(g_{ab}, \nabla_a)$ of a NE-SF null surface $H$.  
They were derived, analyzed and solved in \cite{ABL1}.  

One of the constraints is called the $0$th Low, and it is  satisfied by  the surface gravity
$\kappa^{(\ell)}$ and the  rotation 1-form potential  $\omega^{(\ell)}$ of an arbitrary  non-vanishing $\ell\in\Gamma(L)$, namely
\begin{equation}\label{zeroth}
\partial_a\kappa^{(\ell)} = {\cal L}_{\ell}\omega^{(\ell)}_a. 
\end{equation}
This constraint  has an immediate consequence for the rotation 2-form  $\Omega_{ab}$. To calculate
$\Omega_{ab}$, we can choose any  null vector field in $H$, in particular, we may   use $\ell_o\in\Gamma(L)$ such that  
$${\ell_o}^a\nabla_a\ell_o^b=0.$$
Then,   the $0$th Law implies
\begin{equation}\label{lo}
0= \partial_a\kappa^{(\ell_o)} = {\cal L}_{\ell_o}\omega^{(\ell_o)}_a. 
\end{equation}
Then it is easy to see, that  
\begin{equation}\label{LlOmega}
{\cal L}_{\ell}\Omega_{ab} = 0 = \ell^a \Omega_{ab} 
\end{equation}
holds for $\ell=\ell_o$, and even for every $\ell^a\in L(H)$.
 
The remaining constraints on $(g_{ab}, \nabla_a)$ can be formulated in terms of a given $\ell\in\Gamma(L)$ and the function 
$v:H\rightarrow \mathbb{R}$.
Every value $v_1$ of the function $v$ defines naturally a section of (\ref{Pi1}), 
$$ s_{v_1}:S\rightarrow H . $$ 
We use the section to pullback  to $S$ the co-vectors defined on $H$. To start with, the pullbacks 
\begin{equation}\label{qtildeOmega}  {s_{v_1}^*}{}_{AB}{}^{ab}g_{ab}\ =\ g_{AB}, \ \ \ \ \ \ \ \ \   s_{v_1}^*{}_{AB}{}^{ab}\Omega_{ab} \ =:\ \Omega_{AB}, \end{equation}
are independent of  the value $v_1$ of the function $v$,  of the choice of a function $v$ itself  and of the null vector field  
$\ell\in\Gamma(L)$.   Without the lack of generality one can rescale every $\ell\in\Gamma(L)$ so that 
$$ \kappa^{(\ell)}\ =\ {\rm const}. $$
Hence, we restrict ourselves to that case. 
Next, consider the pullback
\begin{equation}\label{tildeomega} {\omega}_A\ :=\ s_{v_1}^*{}_{A}{}^{a}\omega^{\ell}_a. \end{equation}
The result depends on the choice of $\ell$ and also on the choice of $v$,
unless $\kappa^{(\ell)}=0$.  However, given those choices, ${\omega}_A$ is
independent of the value $v_1$. Of course, the  two pullbacks $\omega_A$ and $\Omega_{AB}$ are related with each other, 
namely 
$$ d{\omega}_{AB}\ =\ {\Omega}_{AB}. $$

Next, we promote the gradient $\partial_av$ to the one form $n_a\in \Gamma(T^*(H))$,
\begin{equation}\label{n1} n_a :=  -\partial_av,  \end{equation} 
invoke  the corresponding tensor $S_{ab}$ (\ref{Sab}) and pull it back  onto $S$  
\begin{equation}\label{tildeS} {S}_{AB}(v)\ :=\ s_v^*{}_{AB}{}^{ab}S_{ab}\end{equation}
(we have dropped the suffix $1$ at $v_1$).
The result is value of $v$ dependent. The dependence   is another constraint implied by (\ref{EE}),
namely
\begin{equation}\label{ddvS} \frac{d}{dv}{S}_{AB}(v)\ =\  
-\kappa^\ell  S_{AB}(v)\ + \nabla_{(A}\omega_{B)}\ +\  \omega_{A}\omega_{B} \ -\ 
\frac{1}{2} R_{AB} \ +\ \frac{1}{2}\Lambda g_{AB} ,
\end{equation}
where $R_{AB}$ is the Ricci tensor of the metric $g_{AB}$. 
 
In conclusion,  the intrinsic  geometry $(g_{ab}, \nabla_a)$  of a  NE-SF null surface in a vacuum spacetime with the cosmological constant $\Lambda$    can be determined by  choosing  $\ell\in\Gamma(L)$ (in other words:  fixing a scale and origin of the geodesic
parameter $v$) such that 
$$\kappa^{(\ell)}\ =\ {\rm const},$$
a section 
$$ s_{v_1}:S\ \rightarrow\ H ,$$ 
and  pullbacks   by $s_{v_1}^*$ onto $S$:  $g_{AB}$, $\omega_A$ (\ref{tildeomega}), and $S_{AB}$ (\ref{tildeS}).  
The constraints  (\ref{lh}, \ref{Llh},  \ref{zeroth}, \ref{ddvS}) provide the degenerate metric tensor $g_{ab}$ and the covariant derivative 
$\nabla_a$ on $H$. 

\subsection{Vacuum isolated null surfaces}\label{gomega}  
Consider a null NE-SF surface  $H$ of intrinsic geometry $(g_{ab}, \nabla_a)$ that admits $\ell\in \Gamma(L)$ non-vanishing on a dense subset of $H$ and  such that 
\begin{equation}\label{LlD}
[{\cal L}_\ell, \nabla_{a}]\ =\ 0 . 
\end{equation}   
\medskip
Notice, that while the condition (\ref{Llh})  is preserved by every transformation $\ell^a \mapsto f\ell^a$ with arbitrary function
 $f$ defined on $H$,  the condition (\ref{LlD})  is generically invariant only for 
 $$f= f_0={\rm const}.$$
 
 The equalities (\ref{omega}) and (\ref{LlD}) imply
 \begin{equation}\label{Llomega}
{\cal L}_{\ell}\omega^{(\ell)}_a = 0,
\end{equation}
hence the $0$th Law  (\ref{zeroth}) takes the following form
\begin{equation}\label{dkappa}
\partial_a\kappa^{(\ell)} = 0,
\end{equation}
meaning that the surface gravity  $\kappa^{(\ell)}$ is constant along (every connected  part) of the null surface $H$. 
A rescaling of a null symmetry generator $\ell$ by a constant results in rescaling  the surface gravity 
\begin{equation}
\kappa^{(f_0\ell)} \ =\ f_0\kappa^{(\ell)} . 
\end{equation}
Therefore,  generically there are  two  cases: either
\begin{equation}
\kappa^{(\ell)}\ \not= \ 0
\end{equation}
or 
\begin{equation}
\kappa^{(\ell)}\ = \ 0. 
\end{equation}
\medskip

\noindent{\bf Definition 2.}  {\it A null  surface  $H$ (\ref{SxR}) is  isolated whenever it is NE-SF and its intrinsic geometry $(g_{ab}, \nabla_a)$ admits 
$\ell\in \Gamma(L)$, non-vanishing on 
$H$ and  such that 
\begin{equation}
[{\cal L}_\ell, \nabla_{a}]\ =\ 0 . 
\end{equation}   
An isolated null  surface is  non-extremal if the surface gravity $\kappa^{(\ell)}$ does not vanish, and
is called extremal if the surface gravity $\kappa^{(\ell)}$ vanishes.} 
\medskip

There exist exceptional  cases, though,  of isolated null surfaces that admit two or more dimensional null symmetry group
\cite{ABL1}. In those cases, an isolated surface is non-extremal with respect to one  $\ell\in\Gamma(L)$ and extremal with respect 
to another  $\ell_o\in\Gamma(L)$.  Because of those special cases, while considering  isolated surfaces we 
also indicate the vector $\ell$  (up to constant rescaling)  that generates a null symmetry (\ref{LlD}).   

In the  non-extremal case the constraint (\ref{ddvS}) allows to express  the tensor ${S}_{AB}$ 
by $g_{AB}$ and  ${\omega}_A$. Indeed, it follows from (\ref{LlD}), that 
$$\frac{d}{dv}{S}_{AB} = 0, $$
hence the constraint (\ref{ddvS}) reads
\begin{equation}\label{Siso}  
  S_{AB}\  =\ \frac{1}{\kappa^{(\ell)}}\left( \nabla_{(A}\omega_{B)}\ +\  \omega_{A}\omega_{B} \ -\ 
\frac{1}{2}R_{AB} \ +\ \frac{1}{2}\Lambda g_{AB} \right).
\end{equation}

In conclusion, in the non-extremal case of isolated null surface $H$ and the infinitesimal generator $\ell\in\Gamma(L)$ of the symmetry,  
the intrinsic geometry  $(g_{ab}, \nabla_a)$  is determined  by the  degenerate metric   $g_{ab}$,   the  rotation 1-form potential 
$\omega^{(\ell)}_a$ and the value of the self acceleration $\kappa^{(\ell)}$. That data is free modulo the constraints (\ref{lh}, \ref{Llh}, \ref{Llomega}, \ref{dkappa}) and the signature $++$ of $g_{ab}$ projected  to $T(H)/L$. 

The rotation $1$-form potential $\omega^{(\ell)}_a$ and the degenerate metric tensor $g_{ab}$ on $H$ can be reconstructed
given a section $s_{v_1}:S\rightarrow H$ and data  on $S$:  $g_{AB}$ and  $\omega_A$ (\ref{tildeomega}).  The $1$-form $\omega_A$
depends on the section. A transformation 
$$ v' \ =\ v-f, \ \ \ \ \ \ \ \ \ \ \ \ \ \ \ \ \ \ \ \ell^a\partial_af\ =\ 0$$
(where $f:S\rightarrow \mathbb{R}$ can be an arbitrary function) induces the respective change of $\omega^{(\ell)}_a$,
in particular
\begin{equation} \label{gauge}\omega'_A\ =\ \omega_A + \kappa^{(\ell)}\partial_A f.\end{equation}

In the extremal isolated horizon case, on the other hand, data freely defined on $S$ is different. The equation (\ref{ddvS}) with 
$\kappa^{(\ell)}=0$ becomes a constraint on $g_{AB}$ and $\omega_{A}$ (\ref{tildeomega}), namely \cite{ABL1}
\begin{equation}\label{EIH}  
 \nabla_{(A}\omega_{B)}\ +\  \omega_{A}\omega_{B} \ -\ 
\frac{1}{2}R_{AB} \ +\ \frac{1}{2}\Lambda g_{AB}  \ =\ 0 .\end{equation}

Notice, that in the extremal case the pull back (\ref{tildeomega}) is independent of the section $s_{v_1}$. 
\medskip

\noindent{\bf Definition 3.} {\it Given a $2$-dimensional surface $S$ endowed with  a Riemannian metric tensor $g_{AB}$
and a $1$-form $\omega_A$, the vacuum extremal isolated horizon equation with a cosmological constant
$\Lambda$ is the equation (\ref{EIH})}.

Solutions to that equation  lead to a classification  of extremal (degenerate) Killing horizons. 
Therefore, the study of its properties and solutions attracted  interest of mathematical relativists \cite{ABL1,LPextremal,JK,CST}.  
Secondly, every solution to that equation determines an exact solution to the Einstein vacuum equations foliated by bifurcated 
Killing horizons whose common part  is an extremal Killing horizon \cite{LPJfol,LSW}. Those spacetimes are called Near Horizon
Geometries and describe neighborhoods of extremal isolated Killing horizons \cite{NHG}.  Because of that the equation itself 
is (most) often called a near horizon geometry equation.  There are still open questions about
 that equation, for example existence of non-axisymmetric solutions on topological  $S_2$.  
We will see below, that somewhat surprisingly, our study of non-extremal isolated horizon will 
lead us to the new results on the extremal isolated horizon equation.  
 
\section{The Weyl tensor of  vacuum  non-extremal isolated null surfaces with a cosmological constant} 

In this section we consider a $3$-dimensional, non-extremal   isolated null surface $H$ introduced above,
endowed with an intrinsic geometry $(g_{ab}, \nabla_a)$ and a null symmetry generator $\ell\in \Gamma(L)$. 
The surface $H$ is contained in a spacetime $M$ whose metric tensor $g_{\mu\nu}$ satisfies the vacuum Einstein equations 
(\ref{EE}) with a (possibly zero) cosmological constant $\Lambda$. 

There exists an extension of the vector field $\ell^a$ to a vector field $t^\mu$ defined in a neighborhood of $H$ in $M$
such that 
\begin{equation}\label{t}
t\ {}_{|_H}\ =\ \ell, \ \ \ \ \ \ \ \ \ {\rm and} \ \ \ \ \ \ \ \ \  {\cal L}_t g_{\mu\nu}\ {}_{|_H}=\ 0 .
\end{equation}
(We construct an example of $t$ below in Sec. \ref{Weyl}.) 
Henceforth, we assume throughout the paper, that every  $t$ like that is also a symmetry of the spacetime  Weyl tensor 
$C^\alpha{}_{\beta\gamma\delta}$ at $H$:

\medskip

\noindent {\bf Assumption 1.} {\it Every vector field $t$ that satisfies (\ref{t}) satisfies also} 
\begin{equation}\label{LtC}
{\cal L}_t C^{\alpha}{}_{\beta\gamma\delta}\ {}_{|_H}=\ 0.
\end{equation} 
\medskip 

That assumption, which we will refer to as the assumption of stationarity to the second order, allows to attribute all the spacetime Weyl tensor components to the intrinsic geometry of $H$.   

\subsection{The complex invariant of  non-extremal isolated null surface}
Consider the data introduced in Sec. \ref{gomega} and defined on the  $2$-dimensional space  $S$    of the null  geodesics  in  $H$,
that is: a Riemannian metric tensor $g_{AB}$ and a differential $1$-form $\omega_A$.  There are two scalar invariants that can be 
constructed from them. One is the Gaussian curvature $K$ (half of the Ricci scalar) of the metric tensor. The second one characterizes 
the rotation $2$-form $\Omega_{AB}$ in terms of the area $2$-form $\eta_{AB}$, 
 \begin{equation} \label{calO}
\Omega_{AB}\ =:\ {\cal O}\,\eta_{AB} .
\end{equation}  
The key role in our paper will be played by a suitable complex valued combination of the  real valued invariants.
\medskip
  
\noindent{\bf Definition 4.} {\it The complex invariant of isolated null surface  $H$  is the complex valued function $\Psi$
defined on the space $S$ of the null  geodesics in  $H$ by the following formula}
\begin{equation} \label{Psi}
\Psi := - \frac{1}{2} \left(K + i{\cal O}\right) .
\end{equation}
\medskip

Both, $K$ and ${\cal O}$ are derived from the freely given $g_{AB}$ and $\omega_A$, respectively. In particular,
the rotation scalar ${\cal O}$, can be expressed by an a priori unconstrained function $U:S\rightarrow \mathbb{R}$,
 \begin{equation} \label{U}
{\cal O}\ =\ - \nabla^A\nabla_A U.
\end{equation}

\subsection{Spacetime null frame adapted to $H$} \label{adaptedframe} 
We will use the Newman-Penrose decomposition of the spacetime Weyl tensor in terms of a spacetime null frame $(e_1{}^\mu, e_2{}^\mu=\overline{e_1{}^\mu}, e_3{}^\mu, e_4{}^\mu)$ and it's 
dual $(e^1{}_\mu,...,e^4{}_\mu)$,  such that
$$ g_{\mu\nu}\ =\ e^1{}_\mu e^2{}_\nu + e^1{}_\nu e^2{}_\mu - e^3{}_\mu e^4{}_\nu + e^3{}_\nu e^4{}_\mu$$

The frame is assumed to be defined in a neighborhood of $H$  in $M$, and to be adapted to the isolated  null surface 
$H$ in a way described below.   To begin with, we set
\begin{equation} \label{l}
e_4   \ {}_{|_H}=\ \ell ,  
\end{equation}
hence the frame is well defined at every  $x\in H$ such that $\ell\not=0$. 
Then, the complex valued vector field $e_1$ is  by definition of a null frame  orthogonal to $\ell$ at $H$, hence, it is tangent to $H$. 
We choose it such that 
\begin{equation} \label{[l,m]}
[\ell, m] = 0,
\end{equation}
where
\begin{equation}\label{m} e_1{}^a  \ {}_{|_H}=:\ m^a.\end{equation}
This choice is possible due to (\ref{Llh}).
Next 
\begin{equation}\label{barm}e_2\ =\ \overline{e_1} . \end{equation}
The outstanding vector field 
$$e_3{}^\mu\ =:\ n^\mu$$ 
is transversal to $H$. 

We restrict further the ambiguities in the vector fields $m^a$ and $n^\mu$ by using the 
$1$-form $n_a$ (\ref{n1}) and assuming about $e^4{}_\mu$, that its pullback $e^4{}_a$ to $H$ coincides with $n_a$, 
\begin{equation} \label{n2} 
e^4{}_a\ {}_{|_H}= n_a .  
\end{equation}
With those frames  $(e_1{}^\mu,...,e_4{}^\mu)$ and the dual $(e^1{}_\mu,...,e^4{}_\mu)$ we  will apply the Newman-Penrose 
formalism \cite{NP} and use the consistent notation\footnote{vectors $k$, $\ell$ introduced in \cite{NP} correspond to vectors $\ell$, $n$ in our work.}. 
\medskip

\noindent{\bf Definition 5.} {\it A null frame $(e_1{}^\mu,...,e_4{}^\mu)$ is called adapted to a given isolated null surface $H$ and 
the null symmetry generator $\ell$ if it satisfies the conditions (\ref{l}, \ref{[l,m]}, \ref{m}, \ref{barm}, \ref{n2}) }.
\medskip

On the other hand, our null spacetime $4$-frame provides a 3-frame tangent to $H$: 
$(m^a ,\bar{m}^a ,\ell^a )$. The corresponding  dual 3-co-frame   $(\bar{m}_a ,{m}_a , -n_a)$  
coincides with the pullback to $H$ of   $(e^1{}_\mu,e^2{}_\mu,e^4{}_\mu)$. Due to (\ref{[l,m]}) the co-frame is Lie dragged 
by $\ell$,
\begin{equation}   {\cal L}_\ell m_a\ = 0\ = {\cal L}_\ell n_a .  \end{equation}

In that  frame, 
$$ g_{ab}\ =\ m_a\bar{m}_b + m_b\bar{m}_a.$$

The components of the intrinsic covariant derivative $\nabla_a$ on  $H$  are expressed below in terms
of the Newman-Penrose coefficients corresponding to the spacetime frame $(e_1{}^\mu,...,e_4{}^\mu)$ introduced above:
\begin{itemize}
\item  $\nabla_a \ell^b\  = \  \left( (\alpha + \bar \beta)m_a + (\bar \alpha +\beta)\bar m_a - (\epsilon +\bar \epsilon)n_a\right) \ell^b$,   \item $\bar m^b  \nabla_an_b\  =\ \lambda m_a + \mu \bar m_a - \pi n_a$;
 \item $m^b  \nabla_a \bar m_b\  =\ -(\alpha - \bar \beta)m_a + (\bar \alpha - \beta) \bar m_a + (\epsilon - \bar \epsilon) n_a  = - \bar m^b  \nabla_a m_b$,
\end{itemize}
 where additional identities hold on $H$ \cite{ABL1}:
 \begin{align}\ 
 \epsilon &= \bar{\epsilon} , \\
 \bar{\mu} &= \mu , \\
\pi &= \alpha + \bar{\beta}\label{pi2} , \\ 
 \kappa^{(\ell)}\ &=\  \epsilon + \bar{\epsilon} .
 \end{align}
Owing to the symmetry (\ref{LlD}), all the coefficients are constant along the null geodesics in $H$.
Consistently with the Newman-Penrose formalism let us denote by $D$, the differential operator $e_4{}^\mu\partial_\mu$
that at $H$ coincides with $\ell^a\partial_a$,
\begin{equation} D\ :=\ \ell^a\partial_a . \end{equation}
Hence, the Newman-Penrose coefficients of $\nabla_a$  satisfy 
\begin{equation}  D\alpha =  D\beta = D\lambda = D\mu = D\epsilon = 0 ,\end{equation} 
and still
$$\kappa^{(\ell)}\ =\ {\rm const}\ \not=\ 0$$   
according to the earlier considerations. 
 
In that co-frame the rotation 1-form potential   takes the following form
\begin{equation}\label{omegaNP}  \omega^{(\ell)}_a\ {=}\ (\alpha + \bar \beta)m_a + (\bar \alpha +\beta)\bar m_a - \kappa^{(\ell)}n_a .
\end{equation} 

It follows that the push forward $\Pi_*m$ onto $S$ is a uniquely  defined vector field on (a neighborhood in)  $S$, 
$$m^A\ :=\  \Pi_*{}^A{}_am^a, $$
and equips $S$ with a null frame for the metric $g_{AB}$,  
$$ g_{AB}\ =\ m_A\bar m_B + \bar m_A m_B  . $$
Using that $2$-frame $(m^A ,\bar{m}^A )$  and the dual 2-co-frame $(\bar{m}_A ,m_A )$, and the Newman-Penrose
coefficients we express the pullbacks  (\ref{tildeomega}, \ref{tildeS}, \ref{Siso}),
\begin{equation}    {\omega}_A =  (\alpha + \bar \beta)m_A + (\bar \alpha +\beta)\bar{m}_A  ,  \end{equation}
\begin{equation}\label{tildeSnp}    {S}_{AB} =  \mu(m_A \bar{m}_B + \bar{m}_A {m}_B) + \lambda {m}_Am_B  + \bar{\lambda} \bar{m}_A \bar{m}_B, 
\end{equation}
where still $S_{AB}$ is determined by $g_{AB}$ and $\omega_A$ by the equality (\ref{Siso}), hence the functions $\lambda$ and $\mu$
can be expressed by $\alpha$, $\beta$  and the differential operators $\delta$ and $\bar{\delta}$, defined consistently with the Newman-Penrose
formalism
 \begin{equation}\delta\ :=\ m^A \partial_A . \end{equation}
 
\subsection{The Weyl tensor}\label{Weyl}

The spacetime  Weyl tensor $C^\mu{}_{\alpha\beta\gamma}$ in the null frame formalism is expressed by five complex valued 
Newman-Penrose components $\Psi_0,...,\Psi_4$,  
\begin{equation}
\Psi_0\ =\ C_{4141}, \ \ \ \ \ \Psi_1\ =\  C_{4341}\ \ \ \ \ \Psi_2\ =\ C_{4123}, \ \ \ \ \ 
\Psi_3\ =\ C_{3432}, \ \ \ \ \ \Psi_4\ =\   C_{3232} .
\end{equation}
 We now use a null frame adapted to $H$ (see the previous subsection) and consider $\Psi_0,...,\Psi_4$ on $H$. 
It turns out, that   the first $4$ components are automatically constant along the null generators
of $H$, that is, that
\begin{equation} \label{ellPsi03}D \Psi_I = 0, \ \ \ \ \ \ \ \ \ \ \ \ \ \ \ \ I=0,1,2,3 . \end{equation}
The general reason for (\ref{ellPsi03}) is,  that $\Psi_0,\,...\,,\Psi_3$ at $H$ are determined by the intrinsic geometry $(g_{ab}, \nabla_a)$ and the adapted null frame, while $\ell$ is an infinitesimal symmetry  of  both, the frame and the intrinsic geometry.  We will provide explicit expressions 
for $\Psi_0,\,...\,,\Psi_3$ below.  

The component $\Psi_4$, on the other hand, is a subject to an evolution equation along the null generators of $H$ with  initial value 
arbitrarily set  at any fixed transversal section $v=v_1$. Specifically, one of the components of the 
tensorial Bianchi identity  
\begin{equation}\label{BianchiC}\nabla_\alpha C^\alpha{}_{\beta\gamma\delta}=0\end{equation}
takes the following form
\begin{equation}\label{ddvPsi4} 0\ =\ D\Psi_4 - \bar\delta \Psi_3 +3\lambda\Psi_2- 2(2\pi+\alpha)\Psi_3 + 2\kappa^{(\ell)} \Psi_4 .
\end{equation}
However, if we  additionally assume that also for $\Psi_4$, 
\begin{equation} \label{ellPsi4} D\Psi_4= 0 \end{equation}
is true, then, for non-zero  $\kappa^{(\ell)}$ the last coefficient  $\Psi_4$ of the Weyl tensor becomes determined by the intrinsic geometry $(g_{ab},\nabla_a)$ of $H$ as well.   
\medskip

\noindent{\bf Assumption 1'.} {\it We are assuming that the Weyl tensor component $\Psi_4$ in a null frame adapted to $H$
 is constant along the null geodesics in $H$. }
 \medskip
 
Technical Assumption 1' follows from the geometrically formulated Assumption 1 and is actually the only reason we needed Assumption 1. 
Let us demonstrate that Assumption 1 implies Assumption 1'.  
We have specified at $H$ the vector field $n^\mu$ that is the $e_3{}^\mu$ element of the
adapted null frame. It is the normalized to $\ell^\mu n_\mu=-1$ null vector  orthogonal to the space-like foliation   of $H$ defined by the constancy
surfaces of the function $v$. Let us extend $n^\mu$ to a neighborhood of $H$ by assuming
$$ n^\mu\nabla_\mu n^\nu=0 .$$
Next, let us   define in a neighborhood of $H$ a vector field $t^\mu$ such that
$$ t^\mu\ {}_{|_H}=\ \ell^a, \ \ \ \ \ \ {\rm and}\ \ \ \ \  {\cal L}_nt\ =\ 0 .$$
By the construction, the resulting vector field $t$ satisfies 
$$ {\cal L}_t e_1{}^\mu\ {}_{|_H}=\   {\cal L}_t e_2{}^\mu\ {}_{|_H}=\ {\cal L}_t e_3{}^\mu\ {}_{|_H}=\ {\cal L}_t e_4{}^\mu\ {}_{|_H}=\ 0$$
hence, also
$${\cal L}_t g_{\mu\nu}\ {}_{|_H}=\ 0 . $$
Now, according to  Assumption 1, 
\begin{equation}\label{LtR}
{\cal L}_t C^{\alpha}{}_{\beta\gamma\delta}\ {}_{|_H}=\ 0 .
\end{equation} 
In particular, that implies that Assumption 1' is satisfied.  It is also easy to show that Assumption 1' implies Assumption 1.

Now, all the Weyl tensor components $\Psi_0, ..., \Psi_4$ can be expressed by $g_{ab}$, $\nabla_a$ and their derivatives with respect to $m^a$ and $\bar{m}^a$.  We derive them below and discuss their properties.  

The components $\Psi_0$ and and $\Psi_1$ vanish identically, due to the vanishing  of the expansion
and shear  of the vector field $\ell$, 
\begin{equation}\label{Psi01} 
\Psi_0 = \Psi_1 = 0 . 
\end{equation}
The Weyl tensor component $\Psi_2$ is determined by the components of the intrinsic covariant derivative $\nabla_a$ on $H$, namely
\begin{equation} \label{eq:psi21}
\Psi_{2} = \bar\delta \beta - \delta \alpha  + \alpha\bar\alpha + \beta\bar\beta - 2\alpha\beta  + \Lambda/6.
\end{equation} 
Not by accident, is it closely related to  the  invariant  
(\ref{Psi}), namely \cite{ABL1}
\begin{equation} \label{Psi2}
\Psi_2 =  \Psi + \frac{\Lambda}{6}, 
\end{equation} 
and obviously is invariant with respect to the allowed transformations of the adapted null frame $(e_1,...,e_4)$. 

The general Newman-Penrose expression for $\Psi_3$ is
\begin{equation} \label{eq:psi3}
\Psi_{3}= \bar\delta\mu - \delta\lambda + \mu(\alpha+\bar\beta)+ \lambda(\bar\alpha- 3\bar\beta) .
\end{equation}
But we remember, that $\lambda$ and $\mu$ can be expressed by $\alpha,\beta,\kappa^{(\ell)}$ via (\ref{Siso}, \ref{tildeSnp}, \ref{pi2}). 
Specifically,
\begin{equation}\label{lambda}
                      \lambda = \frac{1}{\kappa^{(\ell)}} (\bar \delta \pi + \pi(\pi +\alpha - \bar \beta)),
\end{equation}
\begin{equation}\label{mu}
                      \mu = \frac{1}{2 \kappa^{(\ell)}} \bigg(\nabla^A \omega_A + 2\pi \bar \pi - K + \Lambda \bigg),
\end{equation}
where the divergence of $\omega_A$ in terms of Newman-Penrose coefficients reads
\begin{equation}
\nabla^A \omega_A =  \delta\pi + \bar \delta \bar \pi - (\alpha - \bar \beta )\bar \pi - (\bar \alpha - \beta)\pi .
\end{equation}
With the help of the last two equalities,  $\Psi_3$ can be expressed  directly by $\Psi_2$, namely
\begin{equation} \label{Psi3}
\Psi_3\ =\ \frac{1}{\kappa^{(\ell)}}\left(\bar{\delta} + 3\alpha + 3\bar \beta\right)\Psi_2 .
\end{equation}
One can easily see that either by inspection or by applying  
the Bianchi identities (\ref{BianchiC}). The latter  imply
$$0\ =\ D\Psi_3 - \bar{\delta}\Psi_2 + \kappa^{(\ell)}\Psi_3\ - 3\pi\Psi_2 . $$

The last Newman-Penrose component $\Psi_4$ of the Weyl tensor at $H$  by  the assumption (\ref{ellPsi4}) becomes
 \begin{equation} \label{Psi4}
\Psi_{4}= \frac{1}{2\kappa^{(\ell)}}\left(\bar\delta\Psi_{3} - 3\lambda\Psi_{2} + 2(2\pi + \alpha)\Psi_{3}\right) .
\end{equation}

\subsection{The possible Petrov types} 
The conclusion from Sec. \ref{Weyl}  is, that   assuming the vacuum Einstein equations, knowing the value of the cosmological constant $\Lambda$,  and 
making Assumption 1 (or 1'), we can   attribute  the Petrov type of the 
Weyl tensor at  a point $x\in H$  to the  intrinsic geometry $(g_{ab}, \nabla_a)$. The meaning of eqs. (\ref{Psi01}) is that the vector $\ell$ is parallel to 
a double principal null direction of the Weyl tensor. Hence, the Petrov type may be $II, D, III, N$ or $O$.   A necessary condition for the type 
$III$ or $N$ is the vanishing of the component $\Psi_2$ calculated above in eq. (\ref{Psi2}).  Notice, however, that it follows from the formulae (\ref{Psi3}) and (\ref{Psi4}) that for every  open subset of $H$,
$$ \Psi_2\ = 0\ \Rightarrow\ \Psi_3\ =\ \Psi_4\ =0 .$$
In the consequence,  if the Petrov type of the Weyl tensor is constant at $H$,  the only possibilities are: $II, D$ or  $O$.   
If $\Psi_2=0$ at $H$,  then the Petrov type  is necessarily $O$. 
It happens when
\begin{align} \label{KandOmega}
K = \frac{1}{3}\Lambda,\ \ \ \ \ {\rm and}  \ \ \ \ \ \Omega_{AB}=0 . 
\end{align}
Concluding:
\medskip

\noindent{\bf Theorem 0} {\it Suppose  $H$ is a 3-dimensional non-extremal isolated null surface in a 4-dimensional spacetime
such that the vacuum Einstein equations (\ref{EE}) with cosmological constant $\Lambda$ and the Assumption 1 on stationarity to the second order are satisfied. 
If  the Petrov type of the spacetime Weyl tensor is constant on $H$, then the Petrov type is one of the following: II, D, or O.   
In particular, the necessary conditions and sufficient for the  Petrov type to be O on all of $H$ are (\ref{KandOmega}).
 }
\medskip 

If at some point $x\in H$
 $$ \Psi_2(x)\not=0,$$
then the Weyl tensor at $x$  is either of the Petrov type $II$ if
$$ 2\Psi_{3}^2(x) - 3\Psi_{2}(x)\Psi_{4}(x)\ \not=\ 0,$$
or of the Petrov type D when
$$ 2\Psi_{3}^2(x) - 3\Psi_{2}(x)\Psi_{4}(x)\ =\ 0.$$
The (non-)vanishing  of $2\Psi_{3}^2(x) - 3\Psi_{2}(x)\Psi_{4}(x)$ is independent
of the ambiguities remaining in our choice of the null frame $(e_1,e_2,e_3,e_4)$.

\subsection{A remark on the Petrov type of the crossover section}

In this paper we consider isolated null surfaces such that  the infinitesimal generator $\ell$ of
the null symmetry vanishes nowhere. In the non-extremal case, if the integral lines of $\ell$ can be
extended sufficiently far, there is a limit in the spacetime $M$ as we go along each null geodesic
in $H$ in which 
$$\ell\rightarrow 0.$$ 
If 
$$\kappa^{(\ell)}>0,$$
then the limit is in the past, and vice versa. The limiting points form a space-like $2$-surface $S_0\subset M$,
that bounds $H$ from the past or from the future. Let us call it a crossover surface. On the other hand, the Weyl tensor components
$$\Psi_3\ =\ C_{\mu\nu\alpha\delta}n^\mu\ell^\nu n^\alpha{\bar m}^\delta\ = {\rm const}, \ \ \ \ \ \Psi_4\ =\   
C_{\mu\nu\alpha\delta}n^\mu{\bar m}^\nu n^\alpha{\bar m}^\delta={\rm const} $$
 along each line, hence they extend to the limit.  The scaling properties of those components
 as 
 $$ e'_4\ =\ \frac{\ell}{u}, \ \ \ \ \ \ \ \ \  e'_3\ =\  {u}n $$
 are 
 $$ \Psi'_3 \ =  u\Psi_3,  \ \ \ \ \ \ \ \ \  \Psi'_4\ =\  u^2\Psi_4 .$$
So, if we adjust the function $u$ on $H$ such that 
$u\rightarrow 0$ when we approach  $S_0$ while
$e'_4$ and $e'_3$ are both finite in the limit,
then 
$$ \Psi'_3 \ =  0,  \ \ \ \ \ \ \ \ \  \Psi'_4\ =\  0$$
at the crossover surface. Hence,   the null vector  $e'_3$ orthogonal to the
crossover surface is a  double principal null direction of the Weyl tensor. 
That makes the Petrov type of the Weyl tensor  D at the crossover surface.  

In a similar way it is easy to prove that $S'_{AB}\rightarrow 0$ at the crossover surface $S_0$,
that makes the orthogonal null vector $e'_3$ non-expanding and shear free at $S_0$.  

The above concluded properties of the crossover boundary of $H$  are true for every isolated null surface   
extendable  sufficiently far. They do not cause any  constraints on the intrinsic geometry $(g_{ab},\nabla_a)$. 

\section{The Petrov type D isolated null surfaces}
\subsection{Known examples of the Petrov type D isolated null surfaces}
Explicit examples  of the Petrov type D non-extremal isolated null surfaces are provided by the exact solutions to the vacuum Einstein's equations with 
a cosmological constant (positive, negative or zero) that contain non-extremal Killing horizons. A well known family are the non-extremal vacuum black 
hole solutions: 
Schwarzshild and Kerr ($\Lambda=0$), Schwarzshild - de Sitter and   Kerr - de Sitter ($\Lambda>0$), Schwarzshild - anti  de Sitter, Kerr - anti  de Sitter ($\Lambda<0$). Their event horizons,   inner  horizons (Cauchy horizons)  and   cosmological horizons are the   vacuum isolated null surfaces of the Petrov type D.  Another class of  known type D spacetimes that contain non-extremal  isolated horizons are the axi-symmetric  spacetimes foliated by two transversal to each other  families of the  Killing horizons \cite{LPJfol, LSW}.  They are  also known as near horizon geometries \cite{NHG}.  Each of those  Killing horizons is simultaneously extremal and non-extremal,   and of the  
Petrov type $D$. In the case of $\Lambda=0$, it was proven \cite{LPkerr} that every axisymmetric   Petrov type D non-extremal isolated horizon 
is one of the examples listed above. In the current paper we  formulate the generalization of that theorem to $\Lambda\not=0$, (for 
detailed proof see  \cite{DLP2}).

\subsection{The Petrov type D equation}
In this subsection we study the conditions for the Petrov type D  formulated above.
The condition is imposed on the following data freely defined on  a $2$-manifold $S$, the
space of the null generators of an isolated null surface $H$: 
\begin{itemize}
\item $g_{AB}$ - a Riemannian metric tensor,
\item $\Omega_{AB}$ - an exact (rotation) $2$-form, represented by a function  $U$ (\ref{U}),   
\item $\Lambda$ - an arbitrary value of the cosmological  constant.
\end{itemize}

 The surface gravity  
 $$\kappa^{(\ell)}\not=0$$ 
 takes   arbitrarily fixed value  and  disappears from the considerations.

The Petrov type D condition will be expressed by the invariant (\ref{Psi})
\begin{equation} 
\Psi\ =\ \frac{1}{2} \left(\, -K\ +\ i\Delta U \,\right) .
\end{equation}

The metric tensor is represented by  a complex null $2$-co-frame 
$(\bar{m}_A d x^A, m_A d x^A)$,
$$ g_{AB}\ =\ m_A\bar{m}_B\ +\ \bar{m}_A{m}_B .$$
 The  rotation $2$-form  is defined in terms of the rotation $1$-form potential 
$${\omega}_A\ =\ (\alpha+\bar\beta)m_A +  (\bar{\alpha}+\beta)\bar m_A  $$
$${\Omega}_{AB}\ =\ \partial_A{\omega}_B -  \partial_B{\omega}_A. $$
The definition of the Newman-Penrose coefficients $\alpha,\beta:S\rightarrow \mathbb{C}$ 
is completed by the formula for the commutator of the tangent 2-frame $(m^A , 
\bar{m}^A ) $ dual to the coframe, 
\begin{equation} [\delta,\bar{\delta}]\ =\ (\bar{\beta} - \alpha)\delta - (\beta - \bar{\alpha})\bar{\delta} , \end{equation} 
where the vector field $m^A $ is identified with the operator
$$\delta\ =\ m^A\partial_A. $$ 

In terms of $g_{AB}$ and ${\Omega}_{AB}$, the following functions $\Psi_2,\Psi_3$ and $\Psi_4$ are defined on $S$
\begin{align}
\label{Psi2'} \Psi_2 &= \Psi+\frac{\Lambda}{6},  \\  
 \label{Psi3'}\Psi_3 &= \frac{1}{\kappa^{(\ell)}}\left(\bar{\delta} + 3(\alpha+\bar{\beta})\right)\Psi_2,  \\
 \label{Psi4"} \Psi_{4} &= \frac{1}{2\kappa^{(\ell)}}\left(\bar\delta\Psi_{3} - 3\lambda\Psi_{2} + 2(3\alpha+2\bar\beta)\Psi_{3}\right),
\end{align}
with the function $\lambda$ defined in (\ref{lambda}, \ref{pi2}). 

The Petrov type D condition reads 
\begin{align} \label{eq:typeD1'}
\Psi_2(x) \ &\not=\ 0,\\
2\Psi_{3}^2(x) &= 3\Psi_{2}(x)\Psi_{4}(x)\label{eq:typeD1"}.
\end{align}
 After plugging in  the formulae (\ref{Psi2'}, \ref{Psi3'}, \ref{Psi4"})  we obtain an equation on $\Psi_2$
\begin{equation}
 4(\bar\delta \Psi_2)^2\ -\ 3(\alpha-\bar \beta)\Psi_2\bar\delta\Psi_2 - 3\Psi_2 \bar\delta \bar\delta \Psi_2  \ =\ 0 . \end{equation}
Remarkably, the equation can be written in an equivalent  compact form
\begin{equation}\label{typeD}
(\bar\delta + \alpha-\bar\beta) \bar\delta\left(\Psi_2(x)\right)^{-\frac{1}{3}}\ =\ 0 .
\end{equation}
After using     (\ref{Psi2}), the final equation on ($g_{AB}$ and  $\Omega_{AB}$ in terms of) the complex invariant $\Psi$ reads
  \begin{equation}\label{typeDfinal}
  (\bar\delta + \alpha-\bar\beta) \bar\delta\left(\Psi(x)+\frac{\Lambda}{6}\right)^{-\frac{1}{3}}\ =\ 0 .
 \end{equation}
 This is our {\it Petrov type D equation}.       
 
 The operators featuring in (\ref{typeDfinal}) are well known in the GHP formalism \cite{NP}. 
 Using the so-called $edth'$ operator, the equation reads just
 \begin{equation}\label{typeDedth}
\eth' \eth'\left(\Psi_2(x)\right)^{-\frac{1}{3}}\ =\ 0 .
\end{equation} 
 
Notice, that the only dependence on the $1$-form ${\omega}$ is the function $\nabla^A\nabla_A{U}$ present in $\Psi$
that depends on ${\Omega}$ only.    The vector field  $m^A$ (and the corresponding operator  $\delta$)   is defined  modulo 
the local rotations 
\begin{equation}\label{eiphi}
 m'^A = e^{i\varphi}m^A
\end{equation} 
with a real valued function  $\varphi$.  The equation (\ref{typeDfinal}) must be invariant with respect to them. We can see the invariance 
after writing the equation in terms of  the covariant derivative $\nabla_A$ of the metric tensor $g_{AB}$ , namely
\begin{equation}
\left(\bar{\delta} \bar{\delta} -  \left({\bar{m}}^B\nabla_B\bar{m}^A\right)\partial_A\right) \left(\Psi(x)+\frac{\Lambda}{6}\right)^{-\frac{1}{3}}  = 0,
\end{equation}  
that may be written in a more suggestive way,
 \begin{equation}\label{D}
\bar{m}^A\bar{m}^B\nabla_A\nabla_B \left(\Psi(x)+\frac{\Lambda}{6}\right)^{-\frac{1}{3}}  = 0,
\end{equation}   
explicitly invariant with respect to (\ref{eiphi}).

In summary, we have proved the following:
\medskip

\noindent {\bf Theorem 1.} {\it Suppose  $H$ is a 3-dimensional non-extremal isolated null surface in a 4-dimensional spacetime
such that the vacuum Einstein equations (\ref{EE}) with cosmological constant $\Lambda$ and the Assumption 1 on stationarity to the second order (or Assumption 1') 
are satisfied.
Then, the necessary and sufficient condition for the spacetime Weyl  tensor to be  of  the Petrov type D  at each point of the null geodesic  
$x\in S $ is,  that the  invariant $\Psi$ (\ref{Psi}) satisfies the following two conditions: 
 \begin{equation} 
\Psi(x) \  \not= \ -\frac{\Lambda}{6} , 
\end{equation}
 {\it and,}
 \begin{equation}
\bar{m}^A\bar{m}^B\nabla_B\nabla_A\left(\Psi(x)+\frac{\Lambda}{6}\right)^{-\frac{1}{3}}  = 0.
\end{equation}
}
\medskip

A remark is in order. To interpret the Petrov type D equation without using any specific null frame, one can use the concept of the anti-holomorphic covariant derivative.
For every tensor field $T$ defined on $S$, consider the following operation
$$ \nabla^{(0,1)} T:=  \nabla_{\bar{m}} T\otimes {m}_A dx^A ,$$
that turns $T$ into a new tensor. This is the  anti-holomorphic covariant derivative.
It is invariant with respect to the transformations (\ref{eiphi}). With the anti-holomorphic covariant derivative, the  equation (\ref{D}) reads
 \begin{equation}\label{D'}
\left(\nabla^{(0,1)}\right)^2\left(\Psi(x)+\frac{\Lambda}{6}\right)^{-\frac{1}{3}} = 0.
\end{equation}  

\section{The Petrov type D equation and  the near horizon geometry equation}
\subsection{The Petrov type D equation as an integrability condition for the near horizon geometry equation}
Consider a $2$-dimensional manifold $S$, a metric tensor $g_{AB}$, a differential $1$-form ${\omega}_A$ and
a   constant $\Lambda$.  Suppose they satisfy the vacuum extremal isolated horizon equation with a cosmological constant
$\Lambda$
\begin{equation}\label{NHG}  \nabla_{(A}\omega_{B)}\ +\  \omega_{A}\omega_{B} \ -\ 
\frac{1}{2}R_{AB} \ +\ \frac{1}{2}\Lambda g_{AB} \ =\ 0 .
\end{equation}
This equation introduced and studied  in  \cite{ABL1,LPextremal} is better known   as
near horizon geometry equation   \cite{NHG}. 

Our observation is that the corresponding $2$-form 
$$ {\Omega} := d\omega,$$
together with $g_{AB}$ and $\Lambda$ satisfy the  Petrov type D equation (\ref{D}) of the previous subsection.  
The proof is easy, given that the formulae are already at our disposal. Use the given $2$-manifold  $S$ endowed with $g_{AB}$,
${\omega}_A$ and $\Lambda$ to construct on $S\times \mathbb{R}$ the intrinsic geometry of a vacuum non-expanding isolated null surface of
the cosmological constant $\Lambda$. Then, the   near horizon geometry equation (\ref{NHG})
means that the tensor ${S}_{AB}$ (\ref{tildeS},\ref{tildeSnp}) vanishes identically. Indeed,
\begin{equation}\label{Siso'}  
  S_{AB}\  =\ \frac{1}{\kappa^{(\ell)}} \left(\nabla_{(A}\omega_{B)}\ +\  \omega_{A}\omega_{B} \ -\ 
\frac{1}{2}R_{AB} \ +\ \frac{1}{2}\Lambda g_{AB}\right) \ =\ 0. 
\end{equation}
That is the components of ${S}_{AB}$ vanish, namely
$$ \lambda=\mu=0 . $$
In turn, 
\begin{equation} \label{eq:psi3'}
\Psi_{3}\ =\ \bar\delta\mu - \delta\lambda + \mu(\alpha+\bar\beta)+ \lambda(\bar\alpha- 3\bar\beta) \ =\ 0 ,
\end{equation}
and so does $\Psi_4$,
\begin{equation} \label{Psi4'}
\Psi_{4}\ =\ \frac{1}{2\kappa^{(\ell)}}\left(\bar\delta\Psi_{3} - 3\lambda\Psi_{2} + 2(3\alpha+2\bar\beta)\Psi_{3}\right)\ =\ 0 . 
\end{equation}
Hence, the second Petrov type D condition (\ref{eq:typeD1"})  is satisfied.
 The catch is, that   the first condition (\ref{eq:typeD1'}) may be not true in some $x\in S$. The reason is, that in general nothing prevents 
 the invariant $\Psi$ (\ref{Psi}) of $(g_{AB},\omega_A)$ from satisfying 
$$\Psi(x) +\frac{\Lambda}{6}\ =\ 0$$
at some $x\in S$.  

We have thus demonstrated a proof of the following theorem:
\medskip

\noindent{\bf Theorem 2} {\it Suppose a differential $1$-form ${\omega}_A$  and a Riemannian metric tensor 
$$g_{AB}\ = m_A\bar{m}_B\ + \ \bar{m}_A {m}_B,$$ 
both defined on a    $2$-dimensional manifold $S$ satisfy the following  (near horizon geometry) equation:}
\begin{equation}  \nabla_{(A}\omega_{B)}\ +\  \omega_{A}\omega_{B} \ -\ 
\frac{1}{2}R_{AB} \ +\ \frac{1}{2}\Lambda g_{AB} \ =\ 0 ,
\end{equation}
{\it with a constant $\Lambda$. Then, the invariant $\Psi$ defined by  (\ref{Psi}) with}
$${\Omega}_{AB}\ :=\ {\partial_A\omega_B -\partial_B\omega_A}$$
{\it  satisfies the equation}
\begin{equation}
\bar{m}^A\bar{m}^B\nabla_A\nabla_B\left(\Psi(x)+\frac{\Lambda}{6}\right)^{-\frac{1}{3}}  = 0 \end{equation}   
{\it at every $x\in S$ such that }
 \begin{equation}\label{not=}\Psi(x)+\frac{\Lambda}{6} \not= 0.\end{equation}

\medskip
  
 \noindent{\bf Remark}  If  $S=S_2$ (topologically)  and $(g_{AB},\omega_A)$   satisfy the NHG equation (\ref{NHG}), then, by the global 
 argument  given in \cite{LPextremal} (for $\Lambda=0$ but it  can be easily generalized to $\Lambda\not=0$),  the inequality (\ref{not=}) is satisfied
at every $x\in S$, unless $\Psi +\frac{\Lambda}{6}=0$ identically on $S$. The same can be shown with bit more effort for arbitrary orientable compact $S$ (\cite{KL}). 

\medskip

In conclusion, Theorem  2 may also be used   as an integrability condition for the near horizon geometry equation
to investigate  the space of solutions. 
\medskip

\noindent{\bf Remark.}  For every extremal isolated null surface $H$ contained in a vacuum spacetime, the corresponding 
$g_{AB}$, and $\omega_A$ defined on a cross section $S$ of $H$ satisfy the  hypothesis of Theorem 2. Hence, 
they also satisfy the conclusion, that is the type D equation whenever (\ref{not=}).  Nonetheless, the Petrov type of the spacetime Weyl tensor 
at $H$ may be and generically is  different than that of D.  Indeed, Theorem 1 concerns  non-extremal isolated null surface
only.  The technical reason  is that when $\kappa^{(\ell)}=0$,  then an independent of the variable $v$  tensor  $S_{AB}$ 
decouples from $g_{AB}$, and $\omega$, and actually can be arbitrary at a given cross section (\ref{ddvS}).  The same is 
true about $\Psi_4$ (\ref{ddvPsi4}).   The arbitrariness of $S_{AB}$    on $S$ passes to $\Psi_3$ (see (\ref{tildeSnp} ) and (\ref{eq:psi3})).

\subsection{Non-twisting of the  second principal null direction of the Weyl tensor}
At each point of a Petrov type D non-extremal  isolated null surface there are two double principal directions 
of the Weyl tensor. The first one  is orthogonal to $H$ and coincides with the direction of the null symmetry
generator $\ell$.  The second one, however, is generically twisting.  In that generic case  we can not choose 
$e_3$ in our adapted null frame to be pointing in the second double-principal direction because 
in the definition of an adapted null frame we have assumed that both the null vectors $e_3$ and $e_4$ are 
non-twisting at $H$. In this subsection, however,  a second principal null direction  that will emerge below 
will  also be hyper-surface orthogonal due to the extremal isolated horizon (near horizon geometry)
equation  (\ref{NHG}). 
 
Suppose that there is a section of a non-extremal 
isolated null surface $H$ 
$$ s:S\rightarrow H$$
such that  the $1$-form 
$${\omega}_A=s^*\omega^{(\ell)}_\mu$$ 
satisfies the equation (\ref{NHG}). Then, the direct meaning of this equation is that the transversal to $H$ and orthogonal to 
$s(S)\subset H$ null vector field $n^\mu$ is non-expanding and shear-free.  We can also  normalize $n^\mu$ such that 
$$ n_\mu\ell^\mu\ =\ -1$$
and choose a spacetime null frame adapted to $H$, such that 
$$e_3{}^\mu\ {}_{|_H}\ =\ n^\mu .$$
Notice, that if $\ell$ were vanishing at $s(S)$ we could not do that. As we showed in the previous subsection,
in this frame
$$\Psi_3\ {}_{|_H}=\ \Psi_4{}_{|_H}=\ 0 .$$
Therefore,  the isolated null surface $H$ is  of the  Petrov type D.  Moreover, in that case the transversal to
$H$ principal null direction   of the Weyl tensor is orthogonal to the 2-surface $s(S)\subset H$, hence it is
hyper-surface orthogonal.  Applying  to the slice $s(S)$  the symmetry of $H$ generated by $\ell$, we obtain a foliation 
of $H$ whose leaves are $2$-sections of the same properties as 
$s(S)$, hence all $H$ is of the Petrov type D.
  
An opposite implication exists but  is a bit more complicated. Suppose that $H$ is an isolated null surface
of the Petrov type D and that the transversal to $H$ principal null direction of the Weyl tensor is hyper-surface orthogonal. Then, the hyper-surfaces  
are $2$-dimensional, space-like and foliate $H$. Let $s:S\rightarrow H$  be a section such that $s(S)$ is a leaf of the foliation.
We choose in $H$ an adapted null frame $(e_1{}^\mu,...,e_4{}^\mu)$   such that $e_3{}^\mu$ is orthogonal to $s(S)$.  It follows
from the definition of a double principal null direction of the Weyl tensor, that
in this frame
$$\Psi_4\ =\ \Psi_3\ =\ 0 .$$
 The vanishing of $\Psi_3$ and 
$\Psi_4$ on $s(S)$ implies first, that due to (\ref{Psi4})
\begin{equation}  
0=\ \Psi_4\  =\  -\frac{3}{2\kappa^{(\ell)}}\lambda \Psi_2  , 
 \end{equation}
 and since $\Psi_2$ doest not vanish for the type $D$, the shear of the vector field $n^\mu$
 vanishes,
 \begin{equation}  
\lambda \ =\ 0 .
 \end{equation}
 That furnishes the traceless part of the equation (\ref{NHG}). 
 Next, using the eq. (\ref{Psi3}) with $\lambda=0$, 
\begin{equation}  
0=\Psi_{3} \ =\ \bar\delta\mu  + \mu(\alpha+\bar\beta).
\end{equation} 
Now, either
$$ \mu=0,$$
(which in turn implies the trace part of (\ref{NHG})), or the pullback ${\omega}$ of the rotation $1$-form potential 
(\ref{omegaNP}) becomes a pure gradient
\begin{equation}\label{muomega}  
{\omega}_A  \ =\ \partial_A {\rm ln}\mu .
\end{equation}       
As a consequence, the invariant rotation  2-form 
\begin{equation}  
{\Omega}_{AB}  \ =\  0, 
\end{equation}     
which makes $H$ non-rotating.

There is a subtlety about the section orthogonal to the assumed principal null direction
of the Weyl tensor. For $S$ that is not simply connected it may happen, that the section
is not continuous. However, the pullback $(s^*\omega^{(\ell)})_A$ is continuous and differentiable.
That follows from the invariance of the foliation and of $\omega^{(\ell)}_a$ on the flow of
the vector field $\ell^a$.   

We can conclude our findings  in the following way:
\medskip

\noindent{\bf Theorem 3.} {\it Suppose    $H$ is a 3-dimensional non-extremal isolated null surface in a 4-dimensional spacetime
such that the vacuum Einstein equations (\ref{EE}) with cosmological constant $\Lambda$ and  Assumption  1 on stationarity to the second order (or Assumption 1') 
are satisfied.  Let $\ell^a$ be the generator of the null symmetry of $H$. Let $s:S\rightarrow H$  be a  section of (\ref{Pi1}). Then,  

\noindent{$(o)$} The null direction transversal to $H$ and orthogonal to $s(S)$ is non-expanding and
shear-free if and only if the $1$-form 
$${\omega}_A\ :=\ s^*\omega^{(\ell)}_a$$     
satisfies the near horizon geometry equation (\ref{NHG}).
 
\noindent{$(i)$} If the null direction transversal to $H$ and orthogonal to $s(S)$ is non-expanding and
shear-free, then at every $x\in s(S)$ it is  a double principal direction of the Weyl tensor or the Weyl tensor vanishes
at $x$.  

\noindent{$(ii)$} Suppose the rotation scalar 2-form ${\Omega}_{AB}$ does not identically
vanish on $S$. If the null direction orthogonal to $s(S)$ and transversal to $H$ is a double principal null
 direction of the Weyl tensor, then it is non-expanding and shear free.

\noindent{$(iii)$} The symmetry of $H$ generated by $\ell^a$ spreads the slice $s(S)$
to a foliation of $H$ by slices of the same geometric properties. } 
\medskip

Incidentally, in the outstanding non-rotating case, the condition  (\ref{muomega}) imposed on $\mu$ already determined  
via the Bianchi identities by ${\omega}$ becomes a constraint. However, non-trivial solutions do exist: an  
example  is a spherically symmetric section of the Schwarzschild horizon.

\subsection{Isolated null surfaces that are simultaneously non-extremal and extremal}
As we mentioned above it is possible that an  isolated null surface admits two null symmetry 
generators, one  non-extremal and another one extremal. We study this special case now.  
We will argue now, that it is necessarily  of the Petrov type D. 

Consider a non-expanding isolated null surface $H$ of a symmetry generator $\ell$ and  intrinsic geometry $(h,\nabla)$. 
Suppose there exists another  null  vector field tangent to $H$, $\ell_o$ say, non-vanishing on a dense subset of $H$
and such that
 \begin{equation}
\kappa^{(\ell_o)}\ =\ 0, \ \ \ \ \ \ \ \ \ {\rm and} \ \ \ \ \ \ \ \ \   [{\cal L}_{\ell_o}, \nabla_a] \ =\ 0 .  
\end{equation} 
The symmetry generators are related to each other by
$$ \ell = f_1\ell_o, \ \ \ \ \ \ \ \  df_1\ \not= 0.$$
On $H$ we have two rotation $1$-form potentials: $\omega^{(\ell)}$ and $\omega^{(\ell_0)}$.
They are related to each other by
$$ \omega^{(\ell)}\ =\ \omega^{(\ell_o)} \ +\ d{\rm ln}f_1.$$
For every section $s:S\rightarrow H$, we have two pullback $1$-forms: 
$${\omega}\ =\  s^*\omega^{(\ell)}$$
and 
 $${\omega}^{(o)} \ =\ s^*\omega^{(\ell_o)}\ =\ {\omega} - d{\rm ln}s^*f_1 .$$
In the latter one,  ${\omega}^{(o)}$ is independent of the choice  of the section $s$
(because $\kappa^{(\ell_o)}$=0), and  it satisfies the extremal isolated horizon equation 
\begin{equation}
\nabla_{(A}\omega^{(o)}_{B)} \ +\   \omega^{(o)}_{A} \omega^{(o)}_{B}  -\frac{1}{2} R_{AB} \ +\ \frac{1}{2} \Lambda g_{AB} = 0 .
\end{equation}
There exists however a (local) section $s$ such that  
\begin{equation}
\ell_{|_{\tilde S}} =\ell_{o}. 
\end{equation} 
Indeed, there is  the  function $u:H\rightarrow \mathbb{R}$ such that   \cite{ABL1}
\begin{equation}
 Du\ =\ \kappa^{(\ell)}u, \ \ \ \ \ \ \ \ \ \ \ \ \ \ \ \ \ell_o{}^a\partial_a u\ =\ f_2,  \ \ \ \ \ \ \ \ Df_2=0 .
 \end{equation}
The desired section $s$ is defined by the condition  
\begin{equation}
u\ {}_{|_s(S)}=\  \frac{f_2}{\kappa^{(\ell)}} .
 \end{equation}   
For this section $s:S\rightarrow H$ the pullbacks of   $ \omega^{(\ell_o)}$ and   $\omega^{(\ell)}$    coincide, 
$$   \omega^{(o)}\ =\  {\omega},  $$
hence, the equation (\ref{NHG}) holds for ${\omega}$ as well. It implies (see Theorem 2),  
that the Weyl tensor at $H$ is of the Petrove type D. 

The opposite is also true.    Given a non-extremal isolated horizon $H$ of the symmetry generator $\ell$, suppose  there is a section
$s:S\rightarrow H$, such that 
$${\omega}=s^*\omega^{(\ell)}$$
satisfies the extremal isolated null surface  equation (\ref{NHG}).  Consider  the  null vector field $\ell_o$ tangent to $H$ 
defined uniquely by the following two conditions
\begin{equation}
\kappa^{(\ell_o)}\ =\ 0, \ \ \ \ \ \ \ \ \ {\rm and} \ \ \ \ \ \ \ \ \ \ell_o{}_{|_s(S)}\ =\ \ell.  
\end{equation}
It vanishes nowhere, and it follows from  (\ref{NHG}) that \cite{LSW}
\begin{equation}
 [{\cal L}_{\ell_o}, \nabla_a] \ =\ 0 , 
\end{equation}
everywhere on $H$.   Hence, $\ell_o$ makes $H$ an extremal isolated null surface.  That is, $H$ has two null symmetries: the non-extremal one 
$\ell$ and the extremal one $\ell_o$. 
 
We can conclude  this subsection with the following theorem:
\medskip

\noindent{\bf Theorem 4} {\it  Suppose    $H$ is a 3-dimensional non-extremal isolated null surface in a 4-dimensional spacetime
such that the vacuum Einstein equations (\ref{EE}) with cosmological constant $\Lambda$ and  Assumption 1 on stationarity to the second order (or Assumption 1') are 
satisfied;  let $\ell$ be the generator of the null symmetry of $H$. 

\noindent $(i)$ Suppose there exists  a null vector field $\ell_o$ tangent to $H$ and non-vanishing
on $H$,  such that 
\begin{equation}
\kappa^{(\ell_o)}\ =\ 0, \ \ \ \ \ \ \ \ \ {\rm and} \ \ \ \ \ \ \ \ \   [{\cal L}_{\ell_o}, \nabla_a] \ =\ 0 ;  
\end{equation}   
Then, at every point $x\in H$ the spacetime Weyl tensor is either  of the Petrov type $D$
 or of the Petrov type O. 

\noindent $(ii)$ Suppose, there is a section $s:S\rightarrow H$ such that the corresponding ${\omega}$ (\ref{tildeomega}) satisfies
the extremal isolated null surface equation (\ref{NHG}). Then, there is a nowhere vanishing function $f$ defined on $H$ 
such that  the vector field 
$$\ell_o:=f\ell$$ 
satisfies  
\begin{equation}
\kappa^{(\ell_o)}\ =\ 0, \ \ \ \ \ \ \ \ \ {\rm and} \ \ \ \ \ \ \ \ \   [{\cal L}_{\ell_o}, \nabla_A] \ =\ 0.  
\end{equation}   
}     

\section{No hair theorem for the Petrov type D axisymmetric isolated horizons}
In this subsection we consider an axisymmetric non-extremal isolated horizon $H$
such that the space $S$ of the null generators is diffeomorphic to a $2$-dimensional
sphere. We outline the results that will be derived in detail in the accompanying paper 
\cite{DLP2}.  Let  $\Phi^a\in \Gamma(T(H))$ be the generator of  the $1$-dimensional  group of rotations of $H$, 
that preserve the geometry $(g_{ab}, \nabla_a)$ invariant. That is
$${\cal L}_\Phi g_{ab}\ =\ 0, \ \ \ \ \ \ \ \  [{\cal L}_\Phi,\nabla_a]\ =\ 0. $$
In the consequence,  the vector field $\Phi$ Lie drags  the rotation 
$2$-form invariant, 
 $$  {\cal L}_{\Phi}\Omega\ =\ 0 .$$
 The projection $\Pi:H\rightarrow S$ (\ref{Pi1}) 
 pushes forward the vector field $\Phi^a$ onto a uniquely defined vector field ${\Phi}^A$ on $S$,
 $\Phi^A\in \Gamma(T(S))$.
 It becomes a Killing vector of the metric tensor $g_{AB}$, and Lie dragges the pulled back rotation $2$-form
 $$ {\cal L}_{{\Phi}}g_{AB} \ =\ 0,   \ \ \ \ \ \ \ \ \ \ \ \ \ \ \ \  {\cal L}_{{\Phi}}{\Omega_{AB}}\ =\ 0.$$
 Therefore, the problem of finding all the Petrov type D  isolated null surfaces  in this class amounts to solving  
 the Petrov type D equation (\ref{D}) on a $2$-sphere assuming the axial symmetry of the unknowns: a metric $g_{AB}$ 
 and  an exact $2$-form ${\Omega_{AB}}$.   

To characterize axisymmetric  solutions to (\ref{D}) it is convenient to introduce on $S$  a function $\chi$ defined by the area element,
$$ \Phi^A\epsilon_{AB}\ =\ \partial_B\chi.  $$ 
It turns out, that every axisymmetric complex valued function  $\Psi$ that solves   the Petrov type D equation (\ref{D}) has 
a form 
\begin{equation} \Psi\ =\ \frac{1}{(a_1\chi+a_2)^3}\ -\ \frac{\Lambda}{6} , \end{equation}
where $ a_1,a_2\in\mathbb{C}$ are  constant.   

On the $2$-sphere, the complex valued invariant  $\Psi$ satisfies a global topological constraint following
from its relation (\ref{Psi}) with the Gausian curvature $K$ (the Gauss-Bonnet theorem) and the exact $2$-form 
${\Omega}$ (the Stokes' theorem), namely
\begin{equation}\label{intPsi}  \int_S \Psi\,  \epsilon \ =\ -2\pi . \end{equation}
Still, an axisymmetric  metric tensor $g_{AB}$ and rotation $2$-form $\Omega_{AB}$, respectively   integrated out of
$\Psi$ via (\ref{Psi}) have to be regular at the poles.  A detailed analysis shows \cite{DLP2}  (for the 
$\Lambda \ne 0$ case), that the   resulting family of solutions is $2$-dimensional. Moreover, the two parameters
can be chosen to be the area $A$,
$$ A\ =\ \int_S \epsilon$$
 and the angular momentum \cite{ABL2}.  The general formula for the angular momentum of an axisymmetric isolated horizon
reads
\begin{equation}
J \ =\ - \frac{1}{4\pi} \int_{S} \chi\, {\rm Im}\Psi\  \epsilon . 
\end{equation}

In conclusion, the result reads:

\medskip
 
\noindent {\bf Theorem 5} (No-hair) {\it On a topological  $2$-sphere $S$, every axisymmetric solution $g_{AB}$ and ${\Omega}_{AB}$ to the Petrov 
type D equation (\ref{D}) with a cosmological constant $\Lambda$  is uniquely determined by a pair of numbers: the area $A$, and the angular
momentum $J$.  For every pair $(A,J)$ such that:
\begin{itemize}
\item for positive cosmological constant\footnote{the case in which $A=\frac{12\pi}{\Lambda}$ has been excluded.}: $J \in (-\infty,\infty)$ for $A \in \big(0, \frac{ 12 \pi}{\Lambda} \big)$ and $\mid J \mid \ \in \Big[0, \frac{A}{8\pi} \sqrt{\frac{\Lambda A}{12\pi} -1} \Big)$ for $A \in \big(\frac{12\pi}{\Lambda}, \infty \big)$;
\item for negative cosmological constant: $J \in (-\infty,\infty)$ and $A \in (0, \infty)$;
\end{itemize}
there is a unique solution. Every solution gives rise to a non-extremal 
vacuum isolated horizon with the cosmological constant $\Lambda$ and of the Petrov type D.}
\medskip
 
A large class of those isolated horizons  are embeddable in Schwarzschild / Kerr ($\Lambda=0$) or Schwarzschild / Kerr (anti) de 
Sitter (or just de Sitter) spacetimes. There exist also exceptional cases, which admit another null symmetry that is extremal.
Those are embeddable in the near (extremal) horizon spacetimes. 

The spacetime characterization of the Kerr metric is available in the
literature and it provides the uniqueness properties \cite{Mars1, Mars2}. The results outlined in this section were derived for $\Lambda=0$ in \cite{LPkerr}. 
For the generalization to the $\Lambda\not=0$ case see the accompanying paper \cite{DLP2}.

\section{summary} The results of this paper are collected in the Theorems $0-5$.   The vacuum  Einstein Equations (EE) and Assumption 1 on stationarity to the second order at the non-extremal isolated surface $H$  makes the spacetime Weyl tensor $C$ thereon  determined by the intrinsic geometry. In that way properties of $C$   at $H$ become properties of the intrinsic geometry. The components of $C$ are constrained at $H$ in such a way, that   the Petrov type is either II or D or O. The Petrov types III and N can emerge only at measure zero subsets of $H$ (Theorem 0). $H$ can be the Petrov type O only if it is non-rotating and the Gausian curvature of a spacelike section equals $\frac{\Lambda}{3}$.   The condition
for the Petrov type D takes the form of a complex differential equation on certain complex invariant of the intrinsic geometry constructed from the Gausian curvature and the rotation scalar invariant (Theorem 1). In general, the transversal double principal  direction $n'$ of the Weyl tensor is twisting. There are, however,  special cases such that $n'$ is orthogonal to a foliation of $H$ with spacelike slices.  Then, the rotation $1$-form potential  pulled back to any of those slices satisfies the near horizon  geometry equation (Theorem 3).  An important side result is
that the Petrov type D equation is an integrability condition for the  near horizon geometry equation (Theorem 2).  That case has one more geometric characterization, namely  $H$ is both, non-extremal and extremal (Theorem 4).  We also formulate a no-hair theorem valid for the Petrov type D axisymmetric null surfaces of topologically spherical cross-sections (isolated horizons). The intrinsic geometry of each of them is completely determined by two  parameters: the area $A$ and the angular momentum $J$ (Theorem 5). $A$ and $J$ take arbitrary values (of course, $A>0$). A detailed derivation is presented in the accompanying paper \cite{DLP2}. A large class of those isolated horizons is embeddable in Kerr(Schwarzshild) $-$ (anti) de Sitter spacetimes provided $n'$ is twisting. In the hypersurface orthogonal case $H$ can be embedded in a near extremal horizon limit spacetime \cite{Horowitz}. 
 
This work is the first one of the series of papers on
the type D isolated null surfaces   \cite{DLP2, LS, DKLS}.  The
derivation of axisymmetric solutions to the type D equation on
topological sphere - the proof of Theorem 5 above - is presented in
\cite{DLP2}.
The general solution of the type D equation on genus $>0$ sections of
isolated horizons is found in \cite{DKLS}.
The subset  of solutions to the type D equation that is closed with
respect to the map  $(g,\omega) \mapsto (g,-\omega)$  is studied
in \cite{LS}. It is shown that each of those solutions  has an extra
symmetry, that makes it  axisymmetric in the case of topological
sphere. That last result
was motivated by a new work by M.J. Cole, I. Racz  and J.A. Valiente
Kroon \cite{R3} on the Killing spinor characteristic data.
According to that work, in the $\Lambda=0$ case, our type D equation
is the consistency  condition for the Killing spinors characteristic
data. The Killing spinor generates the axial symmetry.

\vspace{1cm}

\noindent{\bf Acknowledgements:}
This work was partially supported by the Polish National Science Centre grant No. 2015/17/B/ST2/02871.

\bibliography{drlp-ih.bib}{}
\bibliographystyle{apsrev4-1}

\end{document}